\documentclass[%
preprint,
 amsmath,amssymb,
 aps,
]{revtex4-1}

\usepackage{graphicx}
\usepackage{bm}
\usepackage{hyperref}

\begin{document}


\title{Principle of Unattainability of Absolute Zero Temperature,\\The Third Law of Thermodynamics,\\ and Projective Quantum Measurements}

\author{Tien D. Kieu}
 \email{tien.d.kieu@gmail.com}
\affiliation{%
Centre for Quantum and Optical Science\\
Swinburne University of Technology, Victoria, Australia
}%




\date{\today}

\begin{abstract}
The Principle of Unattainability rules out the attainment of absolute zero temperature by any finite physical means, no matter how idealised they could be.  Nevertheless, we clarify that the Third Law of Thermodynamics, as defined by Nernst's heat theorem statement, is distinct from the Principle of Unattainability in the sense that the Third Law is mathematically equivalent only to the unattainability of absolute zero temperature by {\em quasi-static adiabatic} processes.  This, on the one hand, leaves open the possibility of attainability of absolute zero by non-adiabatic means, without violating the Third Law.  On the other hand, we point out some apparent incompatibility between the Postulate of Projective Measurement in quantum mechanics and the Principle of Unattainability in that projective measurements of energy could result in zero temperature.

\end{abstract}

\pacs{Valid PACS appear here}
\maketitle





On the one hand, we have the Principle of Unattainability of absolute zero which states that cooling any system to absolute zero temperature in a finite number of steps and within a finite time is physically impossible by any procedure, no matter how idealised the procedure.  This principle has also been strengthened by a recent claim~\cite{Masanes} of its derivation from the laws of quantum mechanics.  On the other hand, we have the Third Law of Thermodynamics originally as statements about the uniqueness of entropy value at absolute zero temperature as approached via different paths~\cite{Wilks}.  This is the version of Nernst's heat theorem, and could also be derived for certain quantum mechanical systems~\cite{Harvey, Aizenman}.  

It is widely and sweepingly claimed that the unattainability and the heat theorem are just two equivalent versions of the same Third Law.  However, there still is an ongoing debate on the relations between the two versions~\cite{Landsberg, Belgiorno1, Belgiorno2}.  Furthermore, the Third Law is not without dispute as there are claims that it could be violated in certain circumstances~\cite{Cleuren, Amikam, Allahverdyan, Sordal}.  It is also speculated that absolute zero temperature could be reached by non-cyclic process~\cite{Atkins}.

In this paper we gather and review some existing proofs in the literature to put forward the arguments that the Third Law of the heat theorem version is not fully equivalent to the Principle of Unattainability.  The heat theorem, mathematically speaking, is only necessary for the latter.  This leaves open the logical possibility of attainability of absolute zero without violating the Third Law so stated.  We also point out herein an intimate connection between the Principle of Unattainability and the von Neumann postulate of projective measurement in quantum mechanics~\cite{vonNeumann}.

\subsection{\label{sec:level1} Zero Temperature and the Ground State}
It should be recognised that at absolute zero temperature only the ground state of a quantum system, with energy bounded from below, is populated and not any of the excited states.  After all, at zero temperature the system has no where to go except to be in its lowest possible energy state.  Conversely, when the system is in its ground state only a temperature of absolute zero can be sensibly and consistently specified/defined for the system.  Having the lowest possible energy, the system cannot yield any energy
but can only accept some inward flow in the form of heat.

That the system being at absolute zero is equivalent to zero occupation of all excited states is further supported by and reflected in the Boltzmann factors, when they are applicable
\begin{eqnarray}
p_i \sim \frac{{\rm population \; of \; the \; excited \; state \; having \; energy} \; E_i}{\rm ground \; state \; population} &=& \exp\{-E_i/kT\}, \label{Boltzmann}
\end{eqnarray}
where $T$ is the system temperature and $k$ is the Boltzmann constant.

This requirement of no occupation of any excited states is quite severe and makes the attainability of absolute zero so challenging.

\subsection{\label{sec:level2} The Third Law and attainability of absolute zero temperature}


The entropy $S(T,\alpha)$ of a system can be expressed as a function of the temperature $T$ and some external parameter $\alpha$
\begin{eqnarray}
S(T,\alpha) &=& S(T=0,\alpha) + \int_0^T \frac{C_\alpha(t)}{t} dt,
\label{entropy}
\end{eqnarray}
where, corresponding to the variable parameter $\alpha$,  $C_\alpha(T)$ is the specific heat and $S(T=0,\alpha)$ is the entropy of the system at absolute zero.  Ernst's heat theorem statement of the Third Law stipulates that at zero temperature the entropy is independent of the variable parameter
\begin{eqnarray}
S(T=0,\alpha) &=& S(T=0,\beta)
\label{Nernst}
\end{eqnarray}
for any $\alpha$ and $\beta$.  Furthermore, the stronger Planck's statement demands that the zero-temperature entropy is also zero
\begin{eqnarray}
S(T=0,\alpha) &=& S(T=0,\beta) = 0.
\label{Planck}
\end{eqnarray}

Ernst's statement is supported by and agrees with Boltzmann's formula for the entropy applied to the system ground state with appropriate degeneracy $W$
\begin{eqnarray}
S &\sim& k\ln W.
\end{eqnarray}
This Boltzmann entropy together with the existence of a ground state, which could be degenerate but still is unique in the sense of being the lowest possible energy level, allow the Third Law to emerge naturally and automatically from quantum mechanics~\cite{Harvey, Aizenman} (for Hamiltonians with spectra bounded from below and their ground states' degeneracy is not dependent on external parameters.)  For a dynamical view of quantum thermodynamics for open quantum systems, see~\cite{Kosloff} and references therein.



As a consequence, the Third Law~(\ref{Nernst}) or~(\ref{Planck}) demands that at zero temperature the adiabat is the same as the isotherm $T=0$~\cite{Pauli3, Callen}.  Then it follows that absolute zero temperature cannot be reached by any adiabatic process because any such a process starting from another adiabat at non-zero $T$ is necessarily different from the adiabat at $T=0$ and thus cannot intersect the adiabat at zero temperature -- different adiabats simply cannot cross.

Recently, the authors of~\cite{Masanes}, see also references therein, presented an interesting study of unitary {\em adiabatic} quantum cooling processes and have been able to quantify a lower bound on the acquired temperature as a function of the cooling time.  Once again and also in the quantum mechanical framework, infinite time is required indeed to arrive at absolute zero temperature by any quasi-static adiabatic quantum process.  


Some heuristic understanding of the quantum scenario is to recognise that quantum adiabatic processes preserve the probability distribution of the populations of the energy levels~\cite{KieuQHE}.  As such, an adiabatic quantum process cannot remove the populations $p_i$~(\ref{Boltzmann}) of excited states, which must exist for any non-zero and however small the initial temperature.
In a reversible adiabatic expansion, for example, the energy level $E_i$ is reduced continuously as the energy gaps become smaller with the expansion; and thus, in order to preserve the probability $p_i$ in~(\ref{Boltzmann}), the temperature must be lowered accordingly, but only continuously.  Nevertheless, absolute zero cannot be obtained as long as $p_i$ is non zero for excited states.  Only until these states merge with the ground state, an asymptotic process taking infinite time in an infinite expansion, when we would have zero temperature.  

To avoid such an infinite expansion and to reduce the probability $p_i$, one could also employ some non-adiabatic process interleaving with the adiabatic expansion above -- such as, for instance, that of isothermal keeping $T$ constant while increasing $E_i$ by compression.  This is the classical text-book approach reviewed in Appendix A.  The probability $p_i$ in~(\ref{Boltzmann}) could then be reduced to zero but once again only {\em asymptotically} -- that is, only in an infinite number of steps -- because  with compression the energy value $E_i$ can only vary continuously, as is generally the case for most processes.

The above heuristic quantum arguments do not rely on whether the degeneracy of the ground state is dependent on some external parameters or not.  That is, in the context of unattainability they are applicable even when neither~(\ref{Nernst}) nor~(\ref{Planck}) is satisfied.



Accordingly and as supported by further classical arguments gathered in Appendix A, the heat theorem version of the Third Law implies the unattainability of absolute zero by any quasi-static adiabatic process.  In Appendix B, we present further arguments for a stronger statement that such a Third Law is indeed {\em equivalent} to, being both mathematically necessary and sufficient for, the unattainability of absolute zero by any quasi-static adiabatic process.  

It is often argued that this should then be sufficient for the equivalence of the Third Law and the Principle of Unattainability (by any process) because it should be able to decompose every process into adiabatic and isothermal process~\cite{Pauli3}.   However, such a decomposition, in fact, is not universal for it is not applicable to all available processes.  A particularly important exception is the measurement process to be recalled in the next section.

\begin{quote}
{\em All of the above thus leaves open the possibility, logically and physically speaking, of attainability of absolute zero, without violating the heat theorem, by non-adiabatic means (which cannot be decomposed into adiabatic and isothermal processes).}
\end{quote}

\subsection{\label{sec:level3} Projective Measurement in Quantum Mechanics and Attainability of Absolute Zero}
In quantum mechanics, besides the unitary dynamical quantum processes that are governed by the Schr\"odinger equation, there are also those of quantum measurements which are non-unitary, non-causal and normally treated as instantaneous.  Quantum measurements are not described by the Schr\"odinger equation and not well understood; there still are many on-going debates and controversies on the problem of quantum measurement.  Nevertheless, quantum measurement is a central concept of the theory of quantum mechanics.

Here, we pay attention to the von Neumann postulate of projective measurements~\cite{vonNeumann}.  Together with unitary evolutions, projective measurements can account for the most general measurements in quantum mechanics.

Let $M$ be a hermitean operator representing an observable, say the energy of a system which has a discrete energy spectrum, with a spectral decomposition
\begin{eqnarray}
M &=& \sum_i E_i P_i,
\end{eqnarray}
where the $E_i$'s are the eigenvalues and
\begin{eqnarray}
P_i &=& |E_i\rangle\langle E_i|,
\end{eqnarray}
is the projection operator, $P_i^2 = P_i$, corresponding to the eigenvector
$|E_i\rangle$ of the observable $M$.  In a postulated projective measurement of the system pure state $|\psi\rangle$ which has a particular eigenvalue $E_i$ as the measured outcome, the state is instantaneously collapsed to the corresponding eigenstate $|E_i\rangle$, that is, projected to the corresponding eigensubspace
\begin{eqnarray}
|\psi\rangle &\longrightarrow& P_i |\psi\rangle /\sqrt{\langle \psi|P_i|\psi\rangle}. \label{reduction}
\end{eqnarray}
The normalisation on the right hand side is to ensure that $\langle \psi|P_i|\psi\rangle$ is the probability for obtaining the particular outcome value $E_i$.  The von Neumann projective measurement is the most ideal measurement but it is mathematically consistent with and has been extensively invoked in order to explain many important and confirmed features and phenomena of quantum mechanics.  

Quantum measurements are non-adiabatic in general and furthermore they reduce the entropies of the measured systems.  Now, given that such a measurement is available, even only as a postulate, we could entertain in principle the situation in which a projective energy measurement forces a system to collapse into its ground state (with some given probability).  Such a scenario would then be a realisation, by non-adiabatic means, of the elusive attainment of absolute zero -- without violating the Third Law as stated in~(\ref{Nernst}), because the projective measurement is not subjected to~(\ref{5}) or~(\ref{II-Law}) of the Second Law of Thermodynamics.

If, on the other hand for whatever reason, absolute zero temperature could not be obtained by any physical process, as ascertained by the Principle of Unattainability, then not all projective measurements could be realised physically.  The von Neumann postulate would then not be tenable.

\subsection{\label{sec:level5} Concluding remarks}
The heat theorem version of the Third Law of Thermodynamics is argued, with the assumption of the Second Law of Thermodynamics, to be mathematically equivalent only to the unattainability of absolute zero by quasi-static adiabatic processes.  This together with the identification of the absolute zero temperature with a system being in its ground state leaves open a logical possibility  to attain the absolute zero  via {\em non-adiabatic, entropy-reducing} processes without violating such a Third Law.


Of those processes, the postulated von Neumann projective measurement, which is also non-unitary and non-causal, offers a theoretical feasibility of collapsing a wave function to its energy ground state, and consequently attaining the absolute zero at the same time.  For simplicity and clarity, we pay attention to Hamiltonians bounded from below and possessing discrete spectra, at least for the lower energy eigenvalues even if the higher lying parts of the spectra could be quasi-continuous -- because at low temperatures we are interested only in the gaps between the ground states and mostly the first few excited states.  

It would be of great significance if an ideal projective measurement could be physically realised leading to zero temperature without destroying the measured system.  Attainment of the absolute zero would have interesting and important consequences not only for thermodynamics but also for quantum information in general, and adiabatic quantum computation in particular.  

A Bose-Einstein condensate in a trap, for example, perhaps with the help of Feshbach resonance might provide a system for a realisation of absolute zero temperature.  To increase the odds of obtaining the ground state, one could first cool the system by adiabatic means down to very low temperature prior to making a projective measurement of the energy.  The cooler the temperature before measuring the better the odds of projecting the system into the ground state.  The duration of the measurement to obtain a definite outcome (sufficient to discriminate the ground state from other excited states) must also be finite.  If it is, such a duration could also be shortened by increasing the energy gap between the first excited and the ground states with more spatial confinement in the trap. 

As another example of a different set up, see~\cite{Matsuzaki} and references therein for a recent proposal for an implementation of projective measurement of energy for an ensemble of qubits.


If, on the other hand, the Principle of Unattainability by any physical means, no matter how idealised it is, always holds true then its upholding would mathematically imply the invalidity of the postulated von Neumann projective measurements -- at least for the projective measurements of energy (of discrete spectra, in a non-destructive manner).  With regard to the ground state, unattainability of absolute zero is sufficient for unattainabilty of the ground state by non-destructive projective measurement.  And in general, unattainability of absolute zero is equivalent to (non-destructive/nondemolition) unattainabilty of the ground state.

Notwithstanding this, all would not be lost even if the projective measurement of the ground state is obtainable {\em destructively}.  In this case, one could first measure projectively and non-destructively the observables compatible with the energy (that is, those observables which commute with the Hamiltonian under consideration) before performing the destructive projective measurement of the energy.  That way, if and when the ground state energy is obtained subsequently, certain properties of the system corresponding to those observables could then be revealed at absolute zero temperature.

The above is also applicable to any pure energy eigenstate, as once obtained a pure state is unitarily accessible to the ground state.

It is prejudicially difficult and uncomfortable to discard the Principle of Unattainability; but to stick with it would mean abandoning or at least suitably amending the cherished postulate of projective measurement, which occupies a critical role in quantum mechanics and has been supported so convincingly, both directly and indirectly, by experimental evidence up to now.  Worse still, quantum mechanics as a theory simply would not survive without an equivalent replacement of the postulate of projective measurement, were the status quo to be abandoned.  Perhaps, a pure state is only an idealisation and with it a projective measurement is also an ideal?

After the completion of this paper, I was informed by the authors of~\cite{Allahverdyan} of their work, in a section of which it was mentioned that the unattainability principle could be violated when the environment is not thermal (eg. microcanoncical systems), and it was then concluded that the unattainability is recovered when taking into account imperfections in preparing the microcanonic state.  

Also  brought to my attention later was the work~\cite{Guryanova}, in which the authors claim that it is impossible to perform ideal projective measurements on quantum systems using finite resources or finite amount of time.  To reach this conclusion, the authors consider quantum measurement model in which the measured system and the measuring pointers are treated as a single combined quantum system subjected to some unitary evolution together.  However, by sticking to such quantum measurement model, the authors could not offer any view or conclusion on the collapse of wavefunctions, whose role is fundamental in both the measurement problem and the transition from quantum to classical.

I am grateful to Peter Hannaford, Adolfo del Campo and the referees for helpful discussions.

\appendix
\section{The Third Law implies Unattainability by adiabatic means}
To show that the Third Law implies the unattainability of absolute zero by adiabatic means, we can argue as follows, see~\cite{Wilks} for example.  In employing an adiabatic process to reduce the system temperature from $T_2$ (with variable parameter $\beta$) to $T_1$ (with variable parameter $\alpha$ and $T_1 < T_2$), we will have in general a non-decreasing in the system entropy,
\begin{eqnarray}
S(T_1, \alpha) &\ge& S(T_2, \beta). \label{5}
\end{eqnarray}
From~(\ref{entropy}) and from the Third Law~(\ref{Nernst},~\ref{Planck}) which demands that the entropy at zero temperature is independent of the variable parameters $\alpha$ and $\beta$, together with the increase of entropy~(\ref{5}) we have
\begin{eqnarray}
\int_0^{T_1} \frac{C_\alpha(t)}{t} dt &\ge& \int_0^{T_2} \frac{C_\beta(t)}{t} dt.
\end{eqnarray}
Were we able to achieve $T_1=0$, then
\begin{eqnarray}
0 &\ge& \int_0^{T_2} \frac{C_\beta(t)}{t} dt,
\end{eqnarray}
which would have been in contradiction with the demand of strict positivity for the specific heat $C_\beta(t)$ for $t>0$.  

\begin{figure}
\begin{center}
\includegraphics{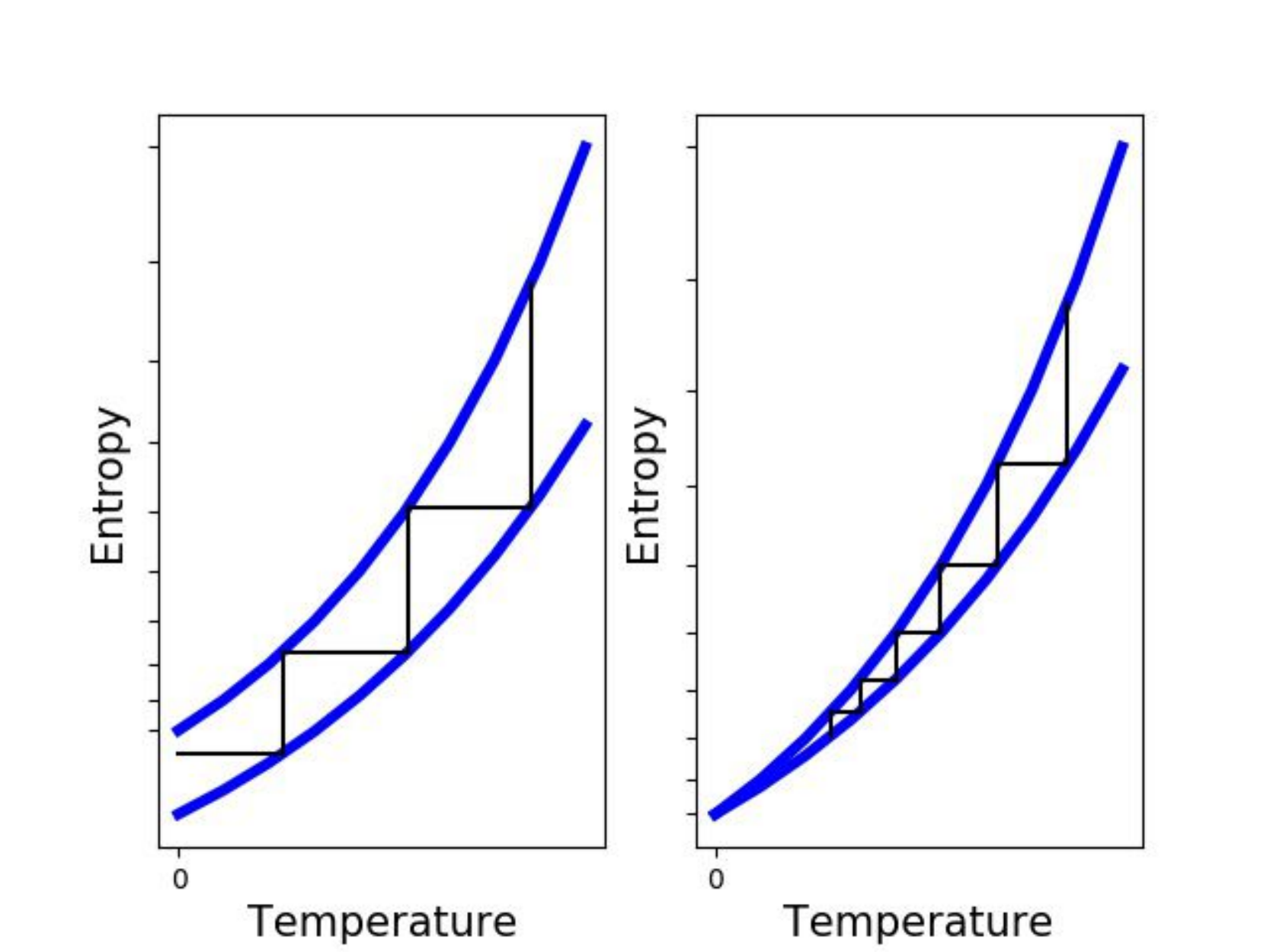}
\caption{
The slopes of the blue curves of constant external parameters X's are non-negative as can be seen from~(\ref{entropy}).  Isothermal processes (vertical segments) successively followed by adiabatic and reversible processes (horizontal segments) could be employed to reduce the system temperature.  While zero temperature could be so reached in a finite number of steps for the case on the left, the Third Law as on the right requires an infinite series of steps.}
\label{fig:figure4}
\end{center}
\end{figure}

Yet another way to reach the same conclusion is depicted in Fig.~\ref{fig:figure4}.  Zero temperature could be reached by a finite series of isothermal processes successively followed by adiabatic (and reversible) processes if the entropy at zero temperature is dependent on the variable parameter $X$, as is the case on the left of Fig.~\ref{fig:figure4}.  But the Third Law demands otherwise, as in the panel on the right, that at zero temperature the entropy has a unique value (zero or not).  It immediately follows that no finite series of isothermal processes (vertical segments) successively followed by adiabatic (and reversible) processes (horizontal segments) can obtain the absolute zero.

Recently, the authors of~\cite{Masanes} have carried out an interesting study of unitary {\em adiabatic} quantum cooling process and been able to quantify a lower bound on the acquired temperature as a function of the cooling time.  Once again and also in the quantum mechanical framework, infinite time is required to arrive at absolute zero temperature by any adiabatic quantum process.

\section{Unattainability by adiabatic means implies the Third Law}
We present here the arguments~\cite{Wilks} that, in the direction opposite to that in the last section, the assumption of unattainability by adiabatic means mathematically implies the Third Law.  

Consider a quasi-static adiabatic change between two states of a system brought about by varying some external parameter from a value $\alpha$ to $\beta$.  As the system adiabatically passes from a state with temperature $T_1$ and entropy $S(T_1,\alpha)$ to another state with temperature $T_2$ and entropy $S(T_2,\beta)$ we then have, by the Second Law of Thermodynamics
\begin{eqnarray}
S(T_1,\alpha) &\le& S(T_2,\beta), \nonumber\\
S(0,\alpha) + \int^{T_1}_0 \frac{C_\alpha(t)}{t} dt &\le& S(0,\beta) + \int^{T_2}_0 \frac{C_\beta(t)}{t} dt. \label{II-Law}
\end{eqnarray}
If $T_2$ is to be zero it follows that
\begin{eqnarray}
\int^{T_1}_0 \frac{C_\alpha(t)}{t} dt &\le& S(0,\beta) - S(0,\alpha). 
\label{b2}
\end{eqnarray}
This is an equation for $T_1$ which will adiabatically lead to an end temperature of absolute zero.  However, if we assert that it is impossible to attain absolute zero from any temperature then the right hand side of~(\ref{b2}) must be non-positive so that~(\ref{b2}), because of the strict positivity of $C_\alpha(t)$ for $t>0$, cannot have any real and non-negative solution for $T_1$.  That is,
\begin{eqnarray}
S(0,\beta) &\le& S(0,\alpha). 
\end{eqnarray}
The same mathematical arguments for the reverse direction to reach temperature $T_1=0$ from an initial temperature $T_2$ will lead to the opposite condition $S(0,\alpha) \le S(0,\beta)$.

Thus from the assumption of Unattainability by adiabatic means we can deduce that
\begin{eqnarray}
S(0,\beta) &=& S(0,\alpha), \forall \alpha, \beta,
\end{eqnarray}
which is precisely Nernst's statement~(\ref{Nernst}) of the Third Law.

Combining the above with the results in the Appendix A,
\begin{quotation}
\em The Third Law of Thermodynamics~(\ref{Nernst}) is mathematically equivalent to the unattainability of absolute zero temperature by any quasi-static adiabatic process.
\end{quotation}


\bibliography{ZeroTemp}

\end{document}